\documentclass[%
reprint,
 superscriptaddress,
 amsmath,amssymb,
 aps,
 floatfix,
 pre,
floatfix,
nolongbibliography,
]{revtex4-2}

\usepackage{comment}
\usepackage{eqnarray}
\usepackage{physics}
\usepackage{dsfont}
\usepackage{tensor}
\usepackage{amsthm}
\usepackage{amsmath}
\usepackage{bbold}
\usepackage{fancyhdr}
\usepackage{subfigure}
\usepackage{epsfig}
\usepackage{bbm}
\usepackage{bm}
\usepackage{tikz}
\usepackage{float}
\usepackage{xcolor}
\usepackage{amsmath}
\usepackage{graphicx}
\usepackage{dcolumn}
\usepackage{bm,braket}
\usepackage{comment}
\usepackage{orcidlink}

\usepackage{soul}

\renewcommand{\selectlanguage}[1]{}
\usepackage{hyperref}

\begin{document}

\preprint{APS/123-QED}

\title{Riding the Wave: Polymers in Time-dependent Nonequilibrium Baths}

\author{Bhavesh Valecha\orcidlink{0009-0000-3219-1786}}
\email[]{bhavesh.valecha@uni-a.de}
\affiliation{Mathematisch-Naturwissenschaftlich-Technische Fakult\"at, Institut f\"ur Physik, Universit\"at Augsburg, Universit\"atsstra{\ss}e 1, 86159 Augsburg, Germany}
\author{Jens-Uwe Sommer\orcidlink{0000-0001-8239-3570}}
\email[]{jens-uwe.sommer@tu-dresden.de}
\affiliation{Leibniz-Institut f\"ur Polymerforschung Dresden, Institut Theorie der Polymere, 01069 Dresden, Germany}
\affiliation{Technische Universit\"at Dresden, Institut f\"ur Theoretische Physik, 01069 Dresden, Germany}
\author{Abhinav Sharma\orcidlink{0000-0002-6436-3826}}
\email[]{abhinav.sharma@uni-a.de}
\affiliation{Mathematisch-Naturwissenschaftlich-Technische Fakult\"at, Institut f\"ur Physik, Universit\"at Augsburg, Universit\"atsstra{\ss}e 1, 86159 Augsburg, Germany}
\affiliation{Leibniz-Institut f\"ur Polymerforschung Dresden, Institut Theorie der Polymere, 01069 Dresden, Germany}

\date{\today}

\begin{abstract}
Directed transport is a characteristic feature of numerous biological systems in response to signals such as nutrient and chemical gradients. These signals are often depend on time owing to the high complexity of interactions in these systems. In this study, we focus on the steady state behavior of polymeric systems responding to such time-dependent signals. We model them as ideal Rouse polymers submerged in a nonequilibrium bath, which is described by a spatially and temporally varying self-propulsion wave field. Through a coarse-graining analysis, we show that these polymers display rich emergent response to the temporal stimuli as a function of their length and topology. In particular, long polymers and structures with ring and star topologies \emph{ride the wave}, displaying a positive drift in the direction of the wave. Whereas, shorter polymers and fully connected structures drift against the wave signal. We confirm these analytical predictions with robust numerical simulations, showing that the response of polymeric systems to temporal stimuli can be controlled by the topology or the length of the polymer.
\end{abstract}

\maketitle

\section{Introduction}
\label{sec:introduction}
Living systems across all scales are characterized by the conversion of stored or ambient energy for various purposes like maintenance of biological processes or to perform systematic movement. These systems are inherently out of equilibrium and have been collectively termed as \emph{active matter}\cite{ramaswamy_mechanics_2010,marchetti_hydrodynamics_2013,ramaswamy_active_2017,vrugt_what_2025}. A ubiquitous class of active systems in biology are polymeric molecules which exhibit directed motion and transport, crucial to many cellular processes\cite{winkler_physics_2020}. Examples are DNA and RNA transcription inside the nucleus\cite{guthold_direct_1999}, ATP-dependent dynamics of chromosomal loci\cite{weber_nonthermal_2012} and chromatin\cite{zidovska_micron-scale_2013} during interphase stage of the cell, and molecular motors determining the structure and conformation of the cellular cytoskeletal network\cite{mackintosh_nonequilibrium_2008, ravichandran_enhanced_2017}. Moreover, many bacteria exist as elongated\cite{auer_bacterial_2019} or chain-like\cite{yaman_emergence_2019} forms and show very rich collective behavior. Mimicking these features in synthetic systems is naturally desirable with applications ranging from drug delivery\cite{wang_nanomicroscale_2012,de_avila_micromotor-enabled_2017,mathesh_supramolecular_2020}, environmental cleanup\cite{jurado-sanchez_micromotors_2018,parmar_micro-_2018} to cargo transport\cite{baraban_transport_2011,ma_catalytic_2015,merlitz_directional_2017}. While this has been achieved in synthetic active systems using external feedback mechanisms\cite{qian_harnessing_2013,mano_optimal_2017,massana-cid_arrested_2021}, it might not be always possible to externally monitor and tune the state of the system.

There have been multiple analytical studies exploring directed motion as an emergent phenomenon in model systems of active particles. In spatially varying activity fields, systems such as chiral and achiral active particle connected to a passive load\cite{valecha_active_2025,vuijk_chemotaxis_2021}, two active particles connected in a dumbbell configuration with fixed orientations\cite{vuijk_active_2022}, and a chiral active dimer\cite{muzzeddu_active_2022} can show accumulation in regions of high activity. These studies have also been extended to include polymeric systems made with active particles that show preferential accumulation depending on the polymer length\cite{muzzeddu_migration_2024, ravichandir_transport_2025}. Polymeric systems are also known to show interesting behavior when immeresed in fluid evironments close to criticality\cite{muzzeddu_structure_2025}. However, treatment of such systems in the more general scenario of temporally and spatially varying activity fields has been limited to single active particles\cite{geiseler_chemotaxis_2016,geiseler_self-polarizing_2017, merlitz_linear_2018} or an active-passive dimer\cite{muzzeddu_taxis_2023}. 

Motivated to fill this gap, we consider active Rouse polymers in contact with a nonequilibrium bath, which appears as a spatially and temporally varying activity wave field experienced by the individual monomers. We show analytically that these polymers show directed transport in response to the activity waves, characterized by a systematic drift. Specifically, long polymer chains display a drift along the wave propagation localizing at the wave crests, whereas short chains drift against the wave. Furthermore, the polymer architecture also affects the response to a traveling wave, with the star topology outperforming other architectures with same number of monomers.
\section{The model \& Results}
\label{sec:the_model_results}
\begin{figure}[t]
    \centering
    \includegraphics[width=9.25cm, height=6.cm]{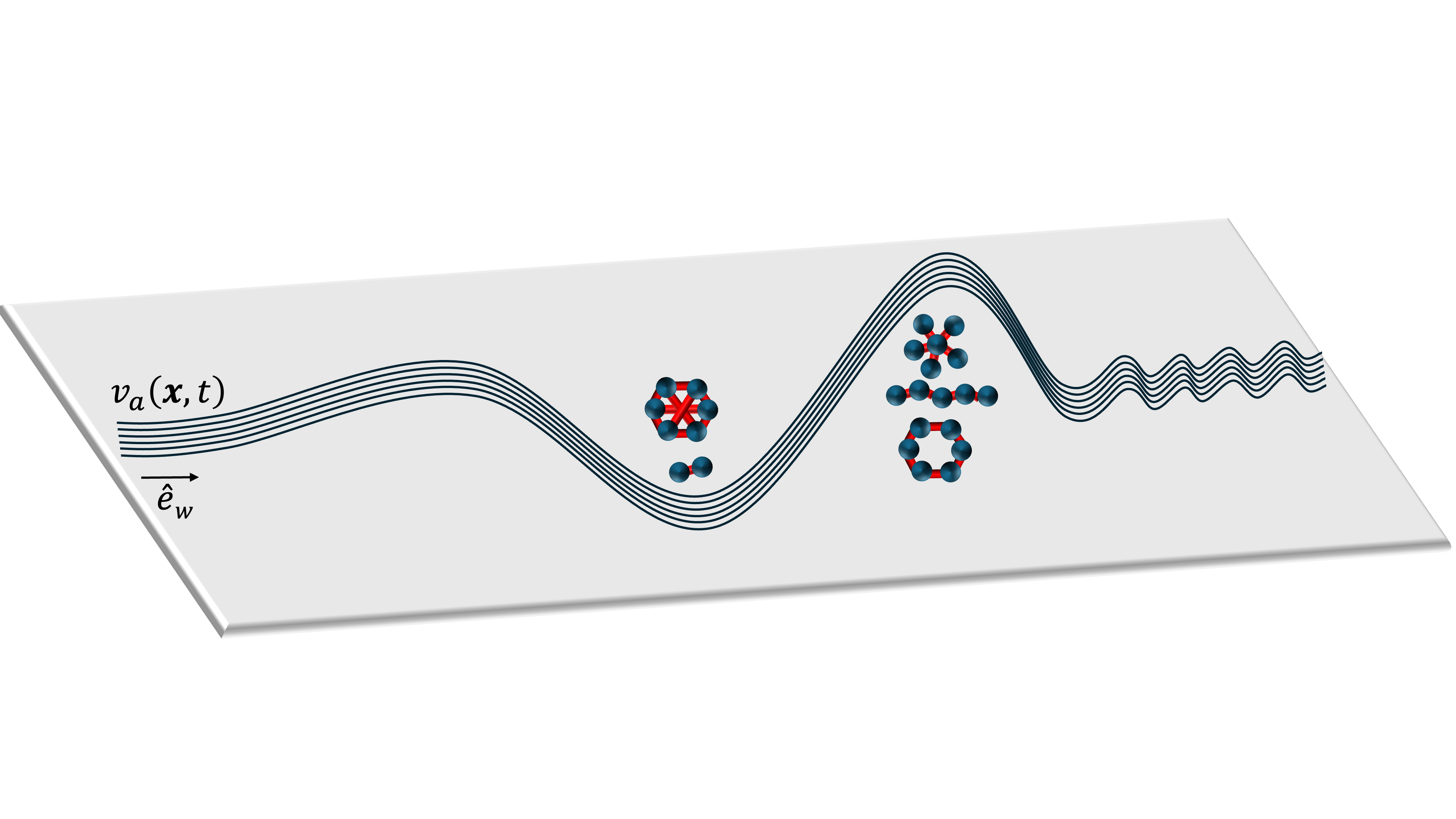}
    \caption{Schematic of active polymers of varying lengths and connectivity experiencing an activity signal, varying in space and time. The signal is represented as a traveling wave $v_a(\boldsymbol{x},t)$ propagating along $\hat{e}_w$. Polymers with a higher degree of polymerization and low connectivity follow the wave peaks and drift in the wave direction, whereas, shorter polymers and fully connected structures follow the wave valleys and drift against the wave.}
\label{fig:schematic}
\end{figure}

We consider an active Rouse polymers immersed in a thermal bath at temperature $T$ in $d$-spatial dimensions, composed of $N$ monomer units interacting via the quadratic Hamiltonian
\begin{equation}
    \mathcal{H} = \frac{\kappa}{2}\sum_{i,j = 0}^{N-1}M_{ij}\boldsymbol{X}_i\cdot\boldsymbol{X}_j,
    \label{eq:polymer_hamiltonian}
\end{equation}
where $\boldsymbol{X}_i$ is the position of the $i$-th monomer and $\kappa$ is the strength of the harmonic interaction. $M_{ij}$ are elements of the connectivity matrix  which determines the polymer topology, defined as $M_{ij} = deg[i]\delta_{ij} - A_{ij}$\cite{sommer_statistics_1995}. Here, $deg[i]$ is the number of bonds arising from monomer $i$ and $A_{ij}$ are elements of the adjacency matrix with $A_{ij} = 1$ if monomers $i$ and $j$ are connected and $A_{ij}=0$ otherwise. Neglecting inertial dynamics, we model the monomers as active Ornstein-Uhlenbeck particles (AOUPs)\cite{caprini_active_2019,martin_statistical_2021,caprini_parental_2022}, described by the following set of $N$ coupled overdamped Langevin equations
\begin{equation}
    \dot{\boldsymbol{X}_i} = -\mu\boldsymbol{\nabla}_{\boldsymbol{X}_i}\mathcal{H} + v_a(\boldsymbol{X}_i - \boldsymbol{v}_wt)\boldsymbol{\eta}_i + \sqrt{2D}\boldsymbol{\xi}_i.
\label{eq:monomer_eom}
\end{equation}
Here, $\mu$ represents the monomer mobility, and $\{\boldsymbol{\xi}_i\}$ are $N$ independent zero mean white Gaussian noises, characterized by the correlations $\langle\boldsymbol{\xi}_i(t)\otimes\boldsymbol{\xi}_j(t')\rangle =\delta_{ij}\boldsymbol{\mathds{1}}\delta(t-t')$ with $\otimes$ denoting the outer product. $D$ is the individual monomer diffusivity which relates to the mobility $\mu$ and the temperature $T$ via the Einstein relation $D = \mu T$, with the Boltzmann constant set to unity $k_B = 1$. Each monomer also experiences a spatially and temporally inhomogeneous self-propulsion wave field, which has the functional form $v_a(\boldsymbol{X}-\boldsymbol{v}_wt)$, i.e. the self-propulsion wave is moving with a constant speed of $v_w$ in the direction $\boldsymbol{\hat{e}}_w$, as shown in Fig.\ref{fig:schematic}. This field will henceforth be referred to as the activity field. The functional dependence of the traveling wave on the spatial coordinate has been left general. The self-propulsion on monomer $i$ acts along the orientation vector $\boldsymbol{\eta}_i$, which evolves as an Ornstein-Uhlenbeck process in $d$ dimensions, with correlation time $\tau$ and variance $d^{-1}$,
\begin{equation}
        \dot{\boldsymbol{\eta}_i} =  -\tau^{-1}\boldsymbol{\eta}_i + \sqrt{2(\tau d)^{-1}}\boldsymbol{\zeta}_i.
\label{eq:orientation_eom}
\end{equation}
 $\{\boldsymbol{\zeta}_i\}$ are independent zero mean white Gaussian noises with correlations $\langle\boldsymbol{\zeta}_i(t)\otimes\boldsymbol{\zeta}_j(t')\rangle = \delta_{ij}\boldsymbol{\mathds{1}}\delta(t-t')$. This implies that the orientation vectors also have zero mean and are exponentially correlated $\langle\boldsymbol{\eta}_i(t)\otimes\boldsymbol{\eta}_i(t')\rangle = d^{-1}\boldsymbol{\mathds{1}}\exp(|t-t'|/\tau)$. The variance has been chosen such that the orientation vectors have the property of unit average modulus $\big\langle||\boldsymbol{\eta}_i^{2}||\big\rangle = 1$ for any dimension $d$.

As is usually done while analyzing Rouse polymers, we perform a coordinate transformation to the Rouse domain\cite{doi_theory_1988} as: $\boldsymbol{\chi}_i = \sum_{j}\varphi_{ij}\boldsymbol{X}_j$. Here, $\varphi_{ij}$ is the matrix that orthogonally diagonalizes the connectivity matrix $M_{ij}$, and with its rows normalized to unity. We further move the zeroth Rouse mode to the co-moving frame of the traveling wave, i.e.,  $\boldsymbol{\chi}_{0} \rightarrow \boldsymbol{\chi}_{0} - N\boldsymbol{v}_wt$. The equations of motion for the Rouse modes are then given by
\begin{equation}
    \begin{aligned}
    \boldsymbol{\dot{\chi}}_0 &= -N\boldsymbol{v}_w -\gamma_0\boldsymbol{\chi}_0 + \sum_{j=0}^{N-1}\varphi_{ij}v_a(\boldsymbol{X}_j - \boldsymbol{v}_wt)\boldsymbol{\eta}_j + \sqrt{2D}\boldsymbol{\xi'}_0, \\
    \boldsymbol{\dot{\chi}}_i &= -\gamma_i\boldsymbol{\chi}_i + \sum_{j=0}^{N-1}\varphi_{ij}v_a(\boldsymbol{X}_j - \boldsymbol{v}_wt)\boldsymbol{\eta}_j + \sqrt{2D}\boldsymbol{\xi'}_i, \quad i\ge 1
    \end{aligned}
\label{eq:rouse_eom}
\end{equation}
where, $\tau_0 = 1/\mu \kappa$ is the monomer relaxation time and   $\gamma_i = \mu\kappa\lambda_i$ are the inverse relaxation timescales of the Rouse modes in the absence of activity, which are determined by the eigenvalues $\{\lambda_i\}$ of the connectivity matrix $M_{ij}$. The noises $\{\boldsymbol{\xi'}_i\}$ are also independent white Gaussian noises with the same statistics as $\{\boldsymbol{\xi}_i\}$.  It is important to point out that the presence of the activity term $\sum_{j=0}^{N-1}\varphi_{ij}v_a(\boldsymbol{X}_j - \boldsymbol{v}_wt)\boldsymbol{\eta}_j$ couples all the Rouse modes, unlike the case of a passive Rouse polymer in equilibrium. A few comments about the Rouse modes and the choice of coordinate transformations are in order here. $\boldsymbol{\chi}_0$ is related to the the center of mass of the polymer\cite{doi_theory_1988} by $\boldsymbol{X}_{\text{COM}} = \boldsymbol{\chi}_0/\sqrt{N}$, where the $\sqrt{N}$ factor comes from the normalization of the matrix $\varphi_{ij}$ to have unit normalized rows. Furthermore, this results in the diffusivities of all Rouse modes to scale with a factor $N$, hence the transformation of the zeroth Rouse mode to the co-moving frame with the same factor.

Since the equations of motion governing the Rouse modes and the orientation vectors are Markovian, the joint probability density $\mathcal{P}(\{\boldsymbol{\chi}\},\{\boldsymbol{\eta}\},t)$ evolves according to the Fokker-Planck equation (FPE)\cite{risken_fokker-planck_1996}
\begin{equation}
      \partial_t \mathcal{P} = \big(\mathcal{L}_0+\mathcal{L}_{\eta}+\mathcal{L}_a\big)\mathcal{P},
\label{eq:FPE}
\end{equation}
where the operators $\mathcal{L}_0$, $\mathcal{L}_{\eta}$ and $\mathcal{L}_a$ give rise to the free diffusion, orientation and activity dynamics respectively, and are given by
\begin{equation}
    \begin{aligned}
        \mathcal{L}_0 &= \sum_{i=0}^{N-1} \boldsymbol{\nabla}_i\cdot\big[\gamma_i\boldsymbol{\chi}_i + D\boldsymbol{\nabla}_i\big] + N\boldsymbol{\nabla}_0\cdot\boldsymbol{v}_w, \\
        \mathcal{L}_{\eta} &= (d\tau)^{-1}\sum_{i=0}^{N-1}\boldsymbol{\tilde{\nabla}}_i\cdot\big[d\boldsymbol{\eta}_i + \boldsymbol{\tilde{\nabla}}_i\big], \\
        \mathcal{L}_a & = \sum_{i=0}^{N-1}\boldsymbol{\nabla}_i\cdot\big[-\sum_{j=0}^{N-1}\varphi_{ij}v_a(\boldsymbol{X}_j - \boldsymbol{v}_wt)\boldsymbol{\eta}_j\big],
    \end{aligned}
\label{eq:FPE_operators}
\end{equation}
where we have used the short-hand notation $\boldsymbol{\nabla}_i\equiv\boldsymbol{\nabla}_{\boldsymbol{\chi}_i}$ and $\boldsymbol{\tilde{\nabla}}_i\equiv\boldsymbol{\nabla}_{\boldsymbol{\eta}_i}$. 

We are interested in the transport properties of the active polymers in the presence of a traveling activity signal. To this end, we attempt a coarse-grained description at the level of the center of mass of the polymer $\boldsymbol{X}_{\text{COM}} = \boldsymbol{\chi}_0/\sqrt{N}$, similar to the analyses in\cite{muzzeddu_migration_2024,vuijk_chemotaxis_2021,muzzeddu_taxis_2023,cates_when_2013}. This is done via a moment-expansion of the joint probability density $\mathcal{P}(\{\boldsymbol{\chi}\},\{\boldsymbol{\eta}\},t)$ in the eigenfunctions of the operator $\mathcal{L}_{\eta}$. Following this, we project the FPE Eq.~\eqref{eq:FPE} onto the respective eigenfunctions to obtain time-evolution equations for physically relevant fields like the position density, average orientation, etc. In particular, the marginalized position probability density, $\rho_0(\boldsymbol{\chi}_0,t) \equiv \int d\boldsymbol{\eta}\int_{h\ne0}d\boldsymbol{\chi}_h\mathcal{P}(\{\boldsymbol{\chi}\},\{\boldsymbol{\eta}\},t)$ evolves as
\begin{equation}
    \begin{split}
    \partial_t\rho_0 = -\boldsymbol{\nabla}_0\cdot\bigg[&-N\boldsymbol{v}_w\rho_0-D\boldsymbol{\nabla}_0\rho_0 \\
    &+ \sum_{j=0}^{N-1}\varphi_{0j}\int_{h\ne0}d\boldsymbol{\chi}_h v_a(\boldsymbol{X}_j-\boldsymbol{v}_w t)\boldsymbol{\sigma}_j\bigg],
    \end{split}
\label{eq:position_density_eom}
\end{equation}
where, $\boldsymbol{\sigma}_j(\{\boldsymbol{\chi}\},t)\equiv\int d\boldsymbol{\eta}\,\boldsymbol{\eta}_j\mathcal{P}(\{\boldsymbol{\chi}\},\{\boldsymbol{\eta}\},t)$ is the conditional average orientation of the $j$-th monomer, given that the polymer configuration is $\{\boldsymbol{\chi}\}$. The $\boldsymbol{\sigma}_j(\{\boldsymbol{\chi}\},t)$ dynamics are given by
\begin{equation}
    \begin{split}
    \partial_t\boldsymbol{\sigma}_j =& -\frac{1}{\tau}\boldsymbol{\sigma}_j + N\boldsymbol{\nabla}_0\cdot\boldsymbol{v}_w\boldsymbol{\sigma}_j+ \sum_{k=0}^{N-1}\boldsymbol{\nabla}_k\cdot\big[\gamma_k\boldsymbol{\chi}_k + D\boldsymbol{\nabla}_k\big]\boldsymbol{\sigma}_j \\
    &- \sum_{k=0}^{N-1}\boldsymbol{\nabla}_k\cdot\big[\frac{\varphi_{kj}}{d}v_a(\boldsymbol{X}_j-\boldsymbol{v}_w t)\varrho\big] + \mathcal{O}(\boldsymbol{\nabla}^{2}),
    \end{split}
\label{eq:orientation_vector_eom}
\end{equation}
where $\varrho$ is the joint position probability density $\varrho\equiv\int d\boldsymbol{\eta}~\mathcal{P}(\{\boldsymbol{\chi}\},\{\boldsymbol{\eta}\},t)$, and we have incorporated the dependencies on higher order conditional moments in $\mathcal{O}(\boldsymbol{\nabla}^2)$. Interestingly, we find an equivalence with the active Brownian polymer model\cite{muzzeddu_migration_2024,ravichandir_transport_2025} by substituting $\tau^{-1} = (d-1)D_r$, where the self-propulsion of the active monomers is modeled as rotational Brownian processes with diffusion coefficient $D_r$. Thus, our coarse-grained analysis remains valid for both these models of activity. Moreover, it has recently been shown that in models for active polymers where the orientation dynamics are independent of the position of the active agents, the steady state behavior of the center of mass is determined only by autocorrelation function of the orientational degrees of freedom, independent of the choice of their microscopic dynamics\cite{dinelli_unifying_2026,dinelli_multiscale_2026}.

For an analytical treatment of the hierarchy of evolution equations obtained from the moment expansion, we employ the \emph{small gradients} approximation and the \emph{adiabatic} approximation. That is, we assume that the gradients in activity $\nabla v_a$ are small compared to the persistence length $l_p = v_a\tau$, as well as, the typical bond length of the polymer $l_b = \sqrt{dT/\kappa}$. This allows us to truncate the expansions in equations of motion Eq.~\eqref{eq:position_density_eom} and Eq.~\eqref{eq:orientation_vector_eom} by neglecting dependencies on higher order gradients. We further point out that the marginalized density $\rho_0(\boldsymbol{\chi}_0,t)$ is the only slow variable in our description as it satisfies a continuity equation $\partial_t\rho_0(\boldsymbol{\chi}_0,t) = -\boldsymbol{\nabla}_0\cdot\boldsymbol{\mathcal{J}}(\boldsymbol{\chi}_0,t)$. And, $\{\boldsymbol{\sigma}_j\}$ and all higher order conditional moments are fast variables, characterized by the presence of sink terms, like $-\tau^{-1}\boldsymbol{\sigma}_j$, in their time evolution. Thus, the fast variables are \emph{quasi-static} at timescales at which the slow variable $\rho_0(\boldsymbol{\chi}_0,t)$ evolves. Together, these approximations allow us to close the infinite hierarchy of evolution equations at the level of $\rho_0(\boldsymbol{\chi}_0,t)$ and $\{\boldsymbol{\sigma}_j\}$. In particular, we obtain a local drift-diffusion equation for the marginalized position probability density $\rho_0(\boldsymbol{\chi}_0,t)$
\begin{equation}
    \partial_t\rho_0 = \boldsymbol{\nabla}_0\cdot\big[\mathcal{V}_{\text{eff}} - \boldsymbol{\nabla}_0(\mathcal{D}_{\text{eff}}\,\rho_0)\big],
\label{eq:drift_diffusion_eq}
\end{equation}
where, we have introduced the effective drift $\mathcal{V}_{\text{eff}}(\boldsymbol{\chi}_0)$ and effective diffusivity $\mathcal{D}_{\text{eff}}(\boldsymbol{\chi}_0)$ given by
\begin{equation}
\begin{aligned}
    \mathcal{V}_{\text{eff}}(\boldsymbol{\chi}_0) &= \bigg(1-\frac{\epsilon}{2}\bigg)\frac{\tau}{d}\boldsymbol{\nabla}_0\,v_a^2(\frac{\boldsymbol{\chi}_0}{\sqrt{N}}) - N\boldsymbol{v}_w, \\
    \mathcal{D}_{\text{eff}}(\boldsymbol{\chi}_0) &= D + \frac{\tau}{d}v_a^2(\frac{\boldsymbol{\chi}_0}{\sqrt{N}}).
\end{aligned}   
\label{eq:effective_drift_diffusivity}
\end{equation}
While this coarse-graining approach and approximations are based on the analysis introduced in \cite{muzzeddu_migration_2024}, we would like to highlight that the time varying activity field introduces new non-trivial terms proportional to the traveling wave velocity $\boldsymbol{v}_w$ in Eq.~\eqref{eq:position_density_eom} and Eq.~\eqref{eq:orientation_vector_eom}. The details of this coarse-graining procedure, the approximations as well as the treatment of the new terms proportional to $\boldsymbol{v}_w$ can be found in the appendices.

Above equations reveal that the effective diffusivity of the center of mass is enhanced by the presence of activity and, the effective drift is proportional to the gradient of effective diffusivity through the \emph{tactic response parameter} $\epsilon$. This parameter depends on the polymer properties, length and topology, through the persistence time $\tau$ of the active forces and the inverse relaxation times $\{\gamma_i\}$ of the Rouse modes, and is given by
\begin{equation}
    \epsilon = 1 - \sum_{i=1}^{N-1}\frac{1}{1+\tau\gamma_i}.
\label{eq:epsilon}
\end{equation}
 Eq.~\eqref{eq:effective_drift_diffusivity} and Eq.~\eqref{eq:epsilon} constitute the central results of this work. Particularly, the parameter $\epsilon$ dictates the response of the active polymer to the time-varying activity signal. For $\epsilon<0$, polymer tends to localize in the traveling wave maxima, while for $\epsilon>0$, the polymer accumulates in the wave minima. This is captured by the steady state density profile, which can be derived by considering a co-moving box of length $L$ with periodic boundary conditions (see Appendix~\ref{sec:appendix_c})
 \begin{equation}
    \rho_0(\boldsymbol{\chi_0}) = \frac{\mathcal{D}^{-1}_{\text{eff}}(\boldsymbol{\chi}_0)\int_{0}^{L} dx \exp{\big\{-\int_{\boldsymbol{\chi}_0}^{\boldsymbol{\chi}_0 + x}dy\frac{\mathcal{V}_{\text{eff}}(y)}{\mathcal{D}_{\text{eff}}(y)}\big\}}}{\int_{0}^{L}du\int_{0}^{L} dx\,\mathcal{D}_{\text{eff}}^{-1}(u)\exp{\big\{-\int_{u}^{u + x}dy\frac{\mathcal{V}_{\text{eff}}(y)}{\mathcal{D}_{\text{eff}}(y)}\big\}}}.
\label{eq:steady_state_density_co_moving_frame}
 \end{equation}
 
 We now show the effect of the polymer length and topology on the steady state accumulation of the polymer in a temporally and spatially inhomogeneous activity field. We assume that the activity field varies sinusoidally in space only along the direction $\hat{e}_w$ with wavelength $\lambda$,
\begin{equation}
    v_a(\chi) = v_0\big(1 + \sin(\chi/\lambda)\big).
\label{eq:activity_form}
\end{equation}
Although chosen for simplicity, we note that any functional form can be generated as a combination of sinusoidal functions, thus keeping this analysis general. We further note that this generality is limited to the regime where the small gradients approximation and the adiabatic approximation are valid, and hence these results might not hold for strongly nonlinear and steep activity landscapes, for e.g. step-like waves.

\begin{figure}[t]
\centering
  \includegraphics[width=9cm, height=7.cm]{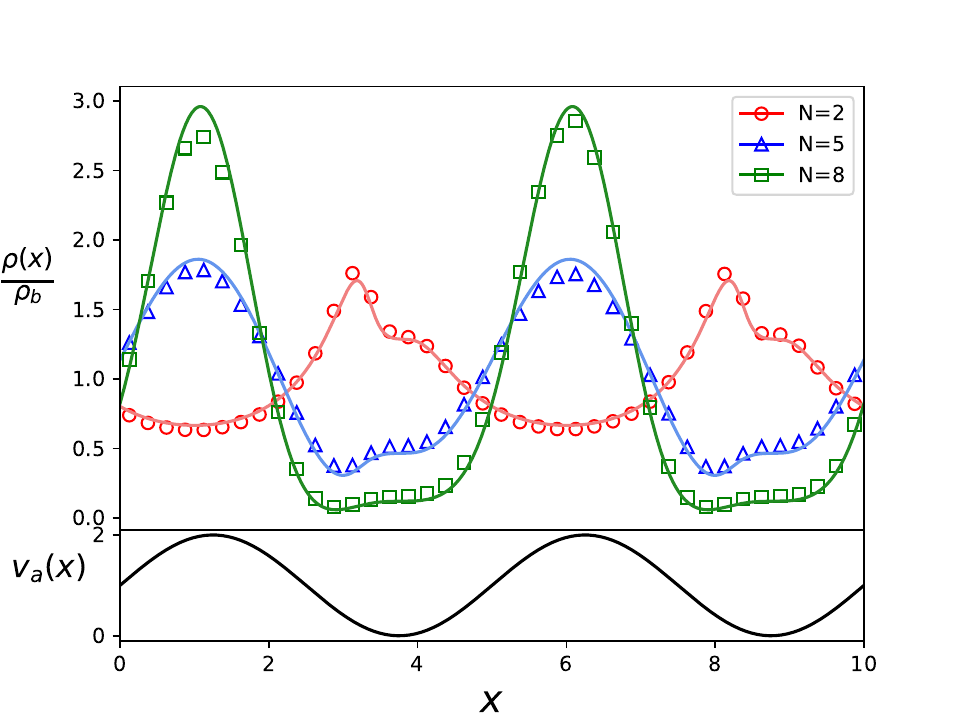}
  \caption{Steady state density profiles for linear active polymers of different lengths in $2$-dimensions. The bottom panel shows the activity field $v_a(x) = v_0\big[1+\sin(4\pi x/L)\big]$ experienced by the active monomers. The $y$-axis in the top panel is normalized with the bulk density, defined as $\rho_b = 1/L$, where $L=10.0$ is the simulation box size with periodic boundary conditions. The parameters of the simulation are $v_w=10^{-2}$, $k_{B}T=10^{-3}$, $\kappa=5.0$, $\gamma=1.0$, $\tau = 0.1$, and $v_0 = 1.0$.}
\label{fig:steady_state_density_waves_length}
\end{figure}

\begin{figure}[t]
\centering
  \includegraphics[width=9cm, height=7.cm]{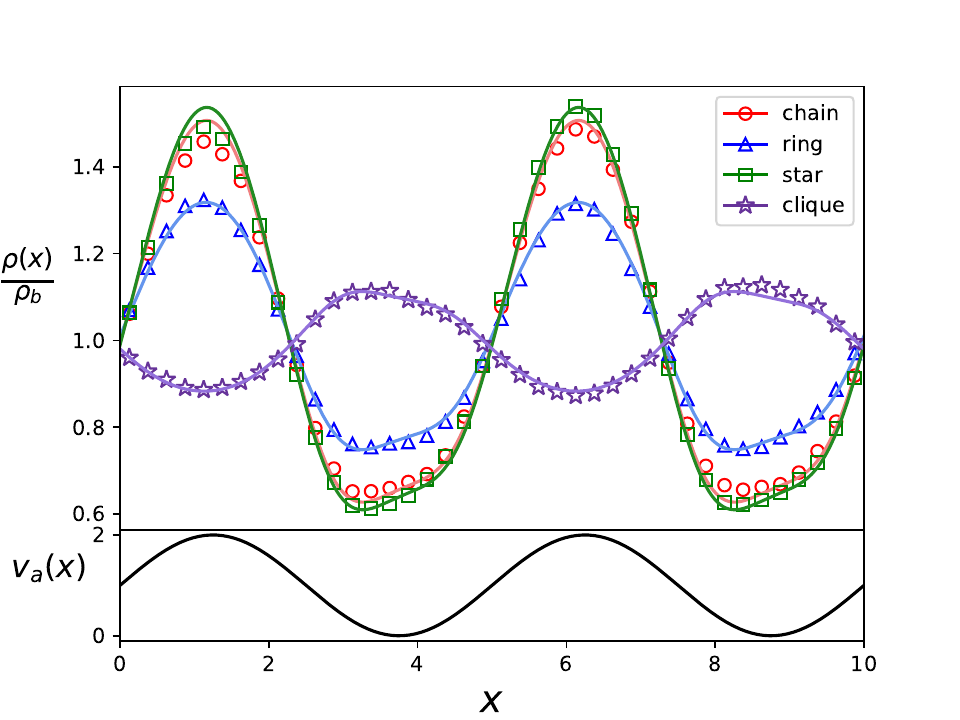}
  \caption{Steady state density profiles for active polymers of different architecture (topology) in $2$-dimensions, each composed of $N=6$ monomers. The bottom panel shows the activity field $v_a(x) = v_0\big[1+\sin(4\pi x/L)\big]$ experienced by the active particle. The $y$-axis in the top panel is normalized with the bulk density, defined as $\rho_b = 1/L$, where $L=10.0$ is the simulation box size with periodic boundary conditions. The parameters of the simulation are $v_w=10^{-2}$, $k_{B}T=10^{-1}$, $\kappa=5.0$, $\gamma=1.0$, $\tau = 0.2$, and $v_0 = 1.0$.}
\label{fig:steady_state_density_waves_topology}
\end{figure}

Fig.~\ref{fig:steady_state_density_waves_length} and Fig.~\ref{fig:steady_state_density_waves_topology} show the steady state density profiles for active polymers with different lengths and architecture, respectively. The solid lines are analytical expressions from Eq.~\eqref{eq:steady_state_density_co_moving_frame}, which are compared with symbols generated with Langevin dynamics simulations (see Appendix~\ref{sec:appendix_d}). These plots corroborate our theoretical prediction that the steady state density can be influenced by two factors: the degree of polymerization and the polymer topology. Specifically, for fixed external parameters $\tau$, $\kappa$ and $\mu$, short polymers tend to localize in regions of low activity and thus follow the wave troughs, whereas, longer polymers accumulate in high activity regions and follow the wave crests. Furthermore, for a polymer with fixed number of monomers, structures with low average connectivity such as a linear chain, a star and a ring tend to follow the wave crests and accumulate where the activity is high, whereas, a fully connected `\emph{clique}' structure shows accumulation in low activity regions and follows the wave troughs.

Moreover, this stationary density is in fact a \emph{non-equilibrium steady state} as it features a non-vanishing constant flux $J_0$. This is reflected in the average drift acquired by the center of the mass of the polymer in the lab frame along the wave direction $\hat{e}_w$, which is given by\cite{goel_stochastic_1974, hanggi_artificial_2009, merlitz_linear_2018}
\begin{equation}
    \begin{aligned}
        v_d &= \sum_{i=0}^{N-1}\frac{\langle\dot{\boldsymbol{X}_i}\rangle}{N} + v_w, \\
        &= \frac{L\big[1-\exp{\big\{-\int_{0}^{L}dy\frac{\mathcal{V}_{\text{eff}}(y)}{\mathcal{D}_{\text{eff}}(y)}\big\}}\big]}{\int_{0}^{L}du\int_{0}^{L} dx\,\mathcal{D}_{\text{eff}}^{-1}(u)\exp{\big\{-\int_{u}^{u + x}dy\frac{\mathcal{V}_{\text{eff}}(y)}{\mathcal{D}_{\text{eff}}(y)}\big\}}}.
    \end{aligned}
\label{eq:average_drift}
\end{equation}

\begin{figure}[t]
\centering
  \includegraphics[width=9.25cm, height=7.25cm]{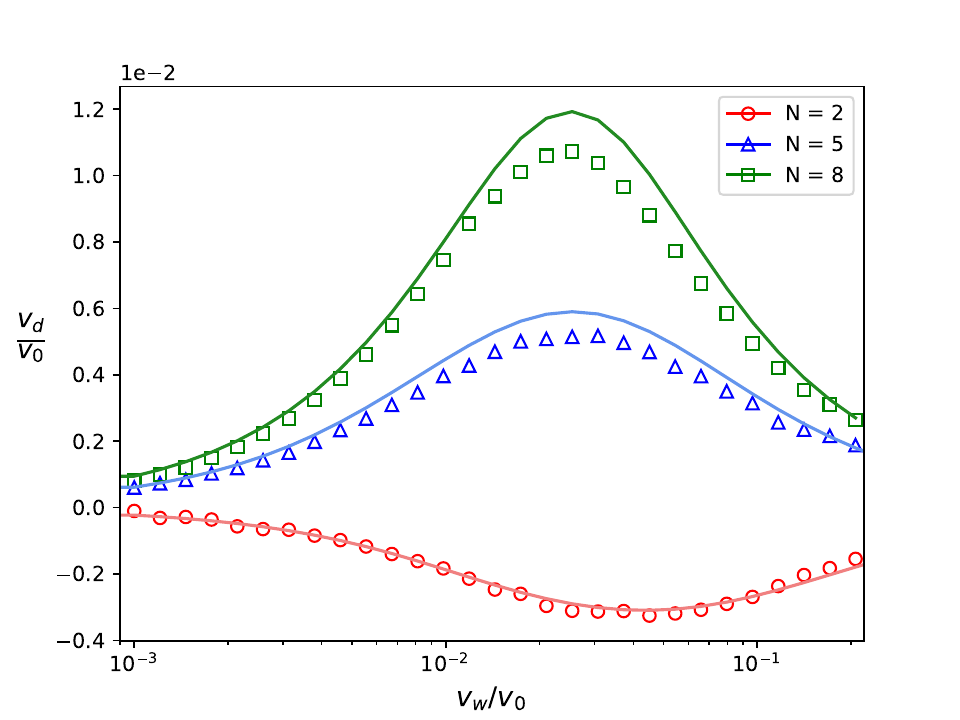}
  \caption{Average drift for different lengths of the linear active polymers in $2$-dimensions. The activity profile is the same as in Fig.~\ref{fig:steady_state_density_waves_length}. Simulation parameters are: $k_{B}T=10^{-2}$, $\kappa=5.0$, $\gamma=1.0$, $\tau = 0.1$, and $v_0 = 1.0$. The results are obtained by running $10^{3}$ independent trajectories and averaging $(\boldsymbol{X}_{\text{COM}}(t) - \boldsymbol{X}_{\text{COM}}(0))/t$ over these trajectories.}
\label{fig:average_drift_length}
\end{figure}

\begin{figure}[t]
\centering
  \includegraphics[width=9.25cm, height=7.25cm]{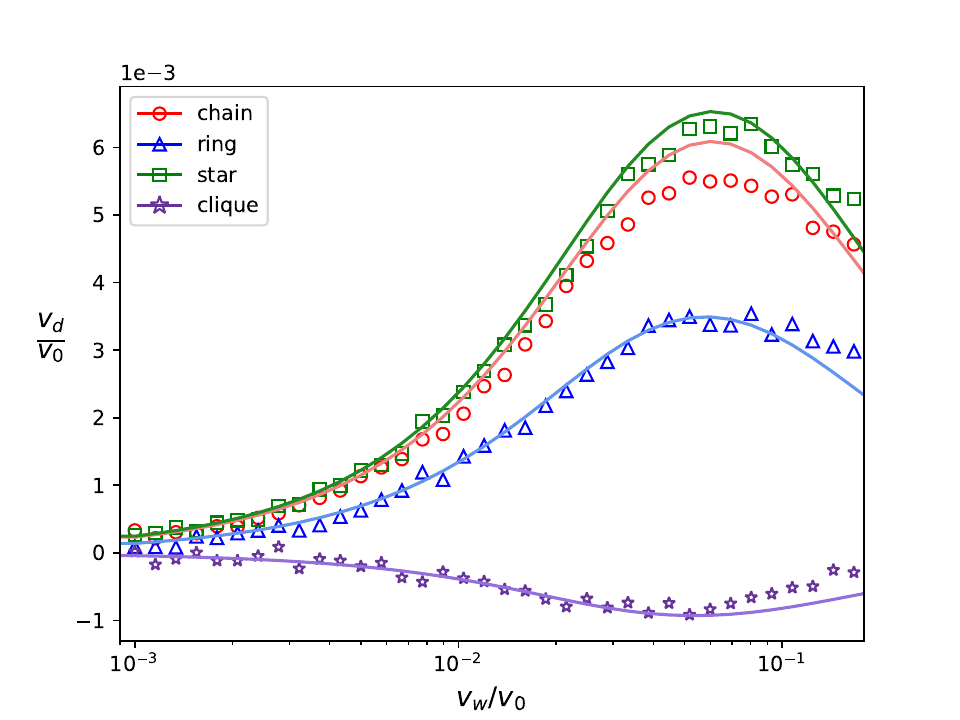}
  \caption{Average drift for different architectures of the active polymers with $N=6$ monomers in $2$-dimensions. The activity profile is the same as in Fig.~\ref{fig:steady_state_density_waves_topology}. Simulation parameters are: $k_{B}T=10^{-1}$, $\kappa=5.0$, $\gamma=1.0$, $\tau = 0.2$, and $v_0 = 1.0$. The results are obtained by running $10^{3}$ independent trajectories and averaging $(\boldsymbol{X}_{\text{COM}}(t) - \boldsymbol{X}_{\text{COM}}(0))/t$ over these trajectories.}
  \label{fig:average_drift_topology}
\end{figure}

\noindent In Fig.~\ref{fig:average_drift_length} and Fig.~\ref{fig:average_drift_topology}, we report the average drift $v_d$ of the center of mass of polymers with varying degrees of polymerization and connectivity, respectively. These plots and Eq.~\eqref{eq:average_drift} show that the drift attained by the polymer depends on the polymer properties, i.e., on $\epsilon$ through the ratio:  $\mathcal{V}_{\text{eff}}/\mathcal{D}_{\text{eff}}$. Specifically depending on the sign of $\epsilon$, the polymer has a non-trivial response to the traveling wave - for $\epsilon < 0$, the polymer displays a positive drift, $v_d>0$, in the direction of the traveling wave, while for $\epsilon>0$, the polymer moves against the wave displaying a negative drift, $v_d<0$. Thus, longer polymers and polymers in linear, star and ring configurations show positive drift, whereas, shorter polymers and a fully connected configuration show negative drift.

These results can be qualitatively explained as follows. From Eq.~\eqref{eq:epsilon}, we see that the sign of $\epsilon$ depends on the contribution of the timescale ratios $(\tau/\tau_i)$  with $\tau_i = \tau_0/\lambda_i=1/\gamma_i$ of individual Rouse modes. Modes with relaxation times longer than the correlation time $\tau$ of active forces, i.e. $\tau/\tau_i \ll 1$, lead to $\epsilon$ being more negative. In fact, when the persistence time is  even much smaller than the monomer relaxation time, i.e. for $\tau/\tau_0 \ll 1$ we get $\epsilon \rightarrow 2-N$, implying that all polymers longer than dimers will follow the traveling wave maxima. For $\tau/\tau_0 > 1$, this threshold is pushed to longer polymers. For linear polymers we have $\lambda_i \simeq i^2/N^2 $, so that the longest relaxation time (the Rouse time $\tau_R$) goes as $\tau_R = \tau_1 \simeq N^2$. Comparing this intrinsic relaxation time with the the persistence time~$\tau$, we can define a crossover chain length $N_c$ for a given set of parameters~$\tau$~and~$\tau_0$, scaling as $N_c \simeq (\tau/\tau_0)^{1/2}$.  Further, for polymers with different topology, we recall that a fully connected structure (clique) we have only one mode which is given by $\tau_1^{\mathrm{clique}} = N \tau_0$. Thus, such highly compact structures have similar tactic response like a single active particle, which always localizes in low activity regions\cite{schnitzer_theory_1993,sharma_brownian_2017}. We conclude that structures with fewer number of connections between monomers leading to the existence of smaller eigenvalues, $\lambda_1$, usually accumulate in high activity regions, and thus drift along with a traveling activity signal. However, we note that there is no simple relation between $\epsilon$ in Eq.(\ref{eq:epsilon}) and the radius of gyration of a Gaussian connected structure, where the latter is given by $R_g^2=l^2 (1/N)\sum_{k=1}^{N-1}1/\lambda_k$ where $l$ denotes the Kuhn length of the macromolecule~\cite{sommer_statistics_1995}.

We reiterate that all the above results hold within the limits of applicability of the the small gradients approximation. This holds for traveling waves with large wavelengths, i.e., $\lambda\gg l_p$, and so when $v_w\ll v_0$. Thus, these theoretical results predict the response of active polymer networks to generic, albeit slow time-dependent activity waves.

\section{Conclusion \& Outlook}
\label{sec:conclusion_outlook}
Time-dependent stimuli are omnipresent in the biological setting across all scales, from cytoskeletal networks in cells to unicellular bacteria. In this work, we have studied the response of a large class of active systems to such stimuli, namely active polymers in traveling activity waves. Using a simple model, we have shown that these systems display non-trivial transport properties in time-varying activity media. In particular, this response crucially depends on the degree of polymerization and the polymer architecture - longer linear polymers and polymers with low average connectivity display a positive drift, accumulating in the wave peaks, in contrast to shorter linear polymers and highly connected polymers, which show a negative drift and accumulate in the wave valleys. This peculiar tactic response can be attributed to the slow Rouse modes of the active polymer that relax on timescales longer than the persistence time of the active forces. We also expect these results to hold for larger polymers ($N\gg1$) as they are known to show strong accumulation in regions of high activity\cite{ravichandir_transport_2025}. These findings extend prior studies of active systems in inhomogeneous activity fields\cite{geiseler_chemotaxis_2016,sharma_brownian_2017,vuijk_chemotaxis_2021,muzzeddu_taxis_2023,muzzeddu_migration_2024} by addressing a previously unexplored regime: active polymer networks subjected to activity landscapes that vary in both space and time. Furthermore, we point out the generality of our approach as it can be easily generalized to consider other interesting systems such as partially active polymer networks\cite{ravichandir_transport_2025} and active polymers with different models of activity\cite{muzzeddu_migration_2024}. 

Recent advances on the experimental front have made possible engineering synthetic active systems with high precision\cite{bechinger_active_2016,lowen_active_2018,niu_modular_2018}. Prime examples include Janus particles, where the self-propulsion arises from diffusiophoresis by catalytic chemical reactions involving hydrogen peroxide\cite{howse_self-motile_2007}, lutidine\cite{buttinoni_dynamical_2013} or hydrazine\cite{gao_catalytic_2014}. Other experimental setups of active colloids also exist, where the self-propulsion is achieved through thermophoresis\cite{jiang_active_2010} or electrophoresis\cite{wurger_thermal_2010}. Some of these synthetic active systems can also be assembled into linear chains and other interconnected structures, composed of - millimeter-sized robots\cite{scholz_surfactants_2021}, magnetic colloidal beads\cite{martinez-pedrero_orientational_2016,mhanna_chain_2022}, silica-beads with magnetic coating in an electric field\cite{di_leonardo_controlled_2016,vutukuri_rational_2017,nishiguchi_flagellar_2018} or diffusiophoretic Janus colloids\cite{biswas_linking_2017}. These platforms could be excellent testbeds for our theoretical predictions. Furthermore, spatio-temporal signaling networks are crucial for various intracellular processes\cite{kholodenko_cell-signalling_2006}. For e.g., temporal dynamics of kinase phosphorylates have been identified to guide the mitotic-spindle self-organization during cell division\cite{kholodenko_cell-signalling_2006,caudron_spatial_2005}. This is a relevant biological scenario where our results could be useful.

Finally, we would like to note that our model does not account for the hydrodynamic interactions between the active polymers and the surrounding solvent, a class of models known as dry active matter models~\cite{marchetti_hydrodynamics_2013}. Furthermore, recent studies have also studied semi-flexible polymers, i.e., polymers with bending interactions (see Supplementary Material of \cite{ravichandir_transport_2025}), as well as, tangentially active polymers, i.e., the self-propulsion is directed along the tangent vector of the polymer backbone\cite{vahid_collective_2025}, in the presence of spatially varying activity landscapes. We defer investigating the extensions of our work in these interesting directions to future studies.

\section{Acknowledgments}
\label{sec:acknowledgements}
A.S.~acknowledges support by the Deutsche Forschungsgemeinschaft (DFG) within Project No.~SH 1275/5-1. J.-U.S.~thanks the cluster of excellence “Physics of Life” at TU Dresden for its support. B.V.~thanks S. Chebolu for her invaluable support, and F. Faedi and A. Pandit for fruitful discussions.
\smallskip
\section{Data Availability}
\label{sec:data_availability}
The data that supports these findings is openly available at\cite{bhavesh_valecha_2026_18835771}.
\bibliography{references}
\clearpage
\appendix
\section{Coarse-graining Procedure}
\label{sec:appendix_a}
Here, we entail the coarse-graining procedure to reach Eq.~\eqref{eq:position_density_eom} and Eq.~\eqref{eq:orientation_vector_eom}, starting from the Fokker-Planck equation Eq.~\eqref{eq:FPE}. This is done in two steps: first by integrating out the orientation degrees of freedom $\{\boldsymbol{\eta}\}$, followed by integrating out all the Rouse modes except $\boldsymbol{\chi}_0$. The moment expansion of $\mathcal{P}(\{\boldsymbol{\chi}\},\{\boldsymbol{\eta}\},t)$ in the eigenfunctions of the operator $\mathcal{L}_{\eta}$ is given by
\begin{equation}
\label{eq:moment_expansion}
    \mathcal{P}(\{\boldsymbol{\chi}\},\{\boldsymbol{\eta}\},t) = \sum_{\boldsymbol{n}}\phi_{\boldsymbol{n}}(\{\boldsymbol{\chi}\},t)u_{\boldsymbol{n}}(\{\boldsymbol{\eta}\}).
\end{equation}
where $\phi_{\boldsymbol{n}}(\{\boldsymbol{\eta}\},t)$ are the expansion modes and $u_{\boldsymbol{n}}(\{\boldsymbol{\eta}\})$ are given in terms of Hermite polynomials in the probabilist's convention\cite{abramowitz_handbook_1965} labeled by $N\times d$ matrix of non-negative integers $\boldsymbol{n}$
\begin{equation}
\label{eq:Ln_eigenfunctions}
    u_{\boldsymbol{n}}(\{\boldsymbol{\eta}\}) = \frac{exp\big\{-\frac{d\sum_j\boldsymbol{\eta}_j^{2}}{2}\big\}\Pi_{i=0}^{N-1}\Pi_{\alpha=1}^{d}H_{n_{i\alpha}}(\sqrt{d}\eta_{i\alpha})}{(2\pi/d)^{Nd/2}}.
\end{equation}
The corresponding eigenfunctions of $\mathcal{L}_{\eta}$ are $\psi_{\boldsymbol{n}} = \tau^{-1}\sum_{i=0}^{N-1}\sum_{\alpha=1}^{d}n_{i\alpha}$. For projecting the FPE onto the eigenfunctions $u_{\boldsymbol{n}}$, we define the auxiliary set of functions $\tilde{u}_{\boldsymbol{n}}(\{\boldsymbol{\eta}\})$, which are orthogonal to $u_{\boldsymbol{n}}(\{\boldsymbol{\eta}\})$ as
\begin{equation}
\label{eq:auxiliary_eigenfunctions}
    \tilde{u}_{\boldsymbol{n}}(\{\boldsymbol{\eta}\}) = \Pi_{i=0}^{N-1}\Pi_{\alpha=1}^{d}\frac{H_{n_{i\alpha}}(\sqrt{d}\eta_{i\alpha})}{n_{i\alpha}!},
\end{equation}
and the orthogonality relation is: $\int\Pi_{i=0}^{N-1}d\boldsymbol{\eta}_i \tilde{u}_{\boldsymbol{m}}u_{\boldsymbol{n}} = \delta_{\boldsymbol{m},\boldsymbol{n}}$. Now, using this orthogonal property, we project the FPE onto the eigenfunctions $\tilde{u}_{\boldsymbol{n}}(\{\boldsymbol{\eta}\})$, and using the properties of the Hermite polynomials, we obtain
\begin{equation}
\label{eq:modes_eom}
    \begin{aligned}
    \partial_t\phi_{\boldsymbol{n}} =&~\psi_{\boldsymbol{n}}\phi_{\boldsymbol{n}} + \mathcal{L}_0\phi_{\boldsymbol{n}} \\
    &- \boldsymbol{\nabla}_i\cdot\varphi_{ij}v_a(\boldsymbol{X}_j-\boldsymbol{v}_wt)\frac{1}{\sqrt{d}}\big[\phi_{b_{j\alpha}\boldsymbol{n}} + (n_{j\alpha}+1)\phi_{b_{j\alpha}^{\dagger}}\big],
    \end{aligned}
\end{equation}
where $~b_{j\alpha}^{\dagger},b_{j\alpha}$ are raising and lowering operators, which increase and decrease the $(i,\alpha)$-component of $\boldsymbol{n}$, respectively. The expansion modes $\phi_{\boldsymbol{n}}$ are proportional to the conditional moments of the orientation vectors $\{\boldsymbol{\eta}\}$ given that the polymer configuration is $\{\boldsymbol{\chi}\}$ as
\begin{equation}
\label{eq:conditonal_moments}
    \begin{aligned}
        \varrho &\equiv \int d\boldsymbol{\eta}~\mathcal{P}(\{\boldsymbol{\chi}\},\{\boldsymbol{\eta}\},t) = \phi_{\boldsymbol{0}}, \\
        \sigma_{i\alpha} &\equiv \int d\boldsymbol{\eta}~\boldsymbol{\eta}~\mathcal{P}(\{\boldsymbol{\chi}\},\{\boldsymbol{\eta}\},t) = \frac{1}{\sqrt{d}}\phi_{b^{\dagger}_{i\alpha}\boldsymbol{0}},
    \end{aligned}
\end{equation}
where $\boldsymbol{0}$ is a $N\times d$ matrix with all elements equal to zero. Now, the time evolution for $\varrho$ and $\sigma_{i\alpha}$ are given by specializing Eq.~\eqref{eq:modes_eom} to $\boldsymbol{n} = \boldsymbol{0}$ and $\boldsymbol{n} = \boldsymbol{b}_{i\alpha}^{\dagger}\boldsymbol{0}$,
\begin{equation}
\label{eq:modes_eom_individual}
    \begin{aligned}
        \partial_t\varrho =& -\sum_{i=1}^{N-1}\boldsymbol{\nabla}_i\cdot\bigg[-\gamma_i\boldsymbol{\chi}_i\varrho - D\boldsymbol{\nabla}_i\varrho\bigg] \\
        &+ \sum_{i,j=0}^{N-1}\varphi_{ij}v(\boldsymbol{X}_j-\boldsymbol{v}_wt)\boldsymbol{\sigma}_j + \boldsymbol{\nabla}_0\cdot[\boldsymbol{v}_w\varrho], \\
        \partial_t\boldsymbol{\sigma}_j =& -\frac{1}{\tau}\boldsymbol{\sigma}_j + N\boldsymbol{\nabla}_0\cdot\boldsymbol{v}_w\boldsymbol{\sigma}_j+ \sum_{k=0}^{N-1}\boldsymbol{\nabla}_k\cdot\big[\gamma_k\boldsymbol{\chi}_k + D\boldsymbol{\nabla}_k\big]\boldsymbol{\sigma}_j \\
    &- \sum_{k=0}^{N-1}\boldsymbol{\nabla}_k\cdot\big[\frac{\varphi_{kj}}{d}v_a(\boldsymbol{X}_j-\boldsymbol{v}_w t)\varrho\big] + \mathcal{O}(\boldsymbol{\nabla}^{2}).
    \end{aligned}
\end{equation}
This constitutes our first coarse-graining step, where we have integrated out the orientation degrees of freedom. We follow this by coarse-graining all Rouse modes except $\boldsymbol{\chi}_0$ from the equation of motion for $\varrho$,
\begin{equation}
\label{eq:position_density_def}
    \rho_0(\boldsymbol{\chi}_0,t) \equiv \int_{h\ne0}d\boldsymbol{\chi}_h~\varrho(\{\boldsymbol{\chi}\},t).
\end{equation}
This gives us the equation of motion for the marginalized position probability density (Eq.~\eqref{eq:position_density_eom}). Similar relations for higher order conditional moments can be derived, however, we recall that our coarse-grained description is limited to the first conditional moment $\{\sigma_{i\alpha}\}$.
\section{Small Gradients Approximation \& Adiabatic Approximation}
\label{sec:appendix_b}
In this appendix, we provide the detailed application of the approximations used to close the infinite hierarchy of equations of motion obtained above. We recall that $\varrho$ and its corresponding coarse-grained version $\rho_0$ satisfy a continuity equation and are the slow variables in our description. This is in contrast to $\{\boldsymbol{\sigma}\}$ and subsequent higher order conditional moments $\phi_{\boldsymbol{n}}$, which are characterized by the presence of decay terms proportional to the persistence time $\tau$ through their corresponding eigenvalues, i.e., $-\psi_{\boldsymbol{n}}\phi_{\boldsymbol{n}}$. This leads to a timescale separation between the slow and the fast modes of our coarse-grained description governed by the correlation time $\tau$ of the active forces. Thus, the timescales at which the slow modes evolve, the fast modes have already relaxed to their \emph{quasi-stationary} value, given by
\begin{equation}
\label{eq:orientation_quasistatic}
    \begin{aligned}
    \boldsymbol{\sigma}_i \approx&~\tau\sum_{i=0}^{N-1}\boldsymbol{\nabla}_i\cdot\big[\gamma_i\boldsymbol{\chi}_i + D\boldsymbol{\nabla}_j\big]\boldsymbol{\sigma}_j +\tau\boldsymbol{\nabla}_0\cdot[\boldsymbol{v}_w\boldsymbol{\sigma}_i]\\
    &- \tau\sum_{j=0}^{N-1}\boldsymbol{\nabla}_j\cdot\bigg[\frac{\varphi_{ji}v_a(\boldsymbol{X}_i-\boldsymbol{v}_wt)\varrho}{d}\bigg] + \mathcal{O}(\boldsymbol{\nabla}^2).
    \end{aligned}
\end{equation}
This is the \emph{adiabatic approximation}. It is important to point out that the validity of this approximation strongly relies on the persistence time $\tau$ of the orientation dynamics. If the persistence time is too long, then there is not enough timescale separation between the slow and fast variables, leading to less accurate predictions using this approximation. 

We will now employ the \emph{small gradients approximation} to arrive at the effective drift-diffusion equation Eq.~\eqref{eq:drift_diffusion_eq}. Specifically, we assume that the variations in the activity field are small compared to the persistence length $l_p = v_a\tau$ and the polymer bond length $l_b = \sqrt{dT/\kappa}$. That is, we only retain terms up to drift and diffusion order in the expression for $\rho_0(\boldsymbol{\chi}_0,t)$, or, only terms which are linear in $\boldsymbol{\nabla}v_a$ in the flux $\boldsymbol{\mathcal{J}}(\boldsymbol{\chi}_0,t)$. We note that this approximation assumes that $\rho_0$ and $\boldsymbol{\mathcal{J}}$ have small variations when the center of mass of the polymer is displaced, and is independent of the internal structure, i.e., the gradients of other Rouse coordinates can be large.

Using the adiabatic expression for $\boldsymbol{\sigma}_i$ in the probability current $\boldsymbol{\mathcal{J}}(\boldsymbol{\chi}_0,t)$, we focus on the contribution coming from the activity term,
\begin{equation}
\label{eq:probility_current}
    \begin{aligned}
    \boldsymbol{\mathcal{J}}_a \equiv& \sum_{j=0}^{N-1}\varphi_{0j}\int_{h\ne0}d\boldsymbol{\chi}_h v_a(\boldsymbol{X}_j-\boldsymbol{v}_w t)\boldsymbol{\sigma}_j, \\
    =&~\tau\sum_{j=0}^{N-1}\varphi_{0j}\int_{h\ne0}d\boldsymbol{\chi}_h v_a(\boldsymbol{X}_j-\boldsymbol{v}_w t)\bigg\{\boldsymbol{\nabla}_0\cdot[\boldsymbol{v}_w\boldsymbol{\sigma}_j]\\
    &+\sum_{i=0}^{N-1}\boldsymbol{\nabla}_i\cdot\big[\gamma_i\boldsymbol{\chi}_i + D\boldsymbol{\nabla}_j\big]\boldsymbol{\sigma}_j \\
    &- \sum_{l=0}^{N-1}\boldsymbol{\nabla}_l\cdot\bigg[\frac{\varphi_{lj}v_a(\boldsymbol{X}_j-\boldsymbol{v}_wt)\varrho}{d}\bigg]\bigg\}+ \mathcal{O}(\boldsymbol{\nabla}^2).
    \end{aligned}
\end{equation}
We will now check the order of gradient in the contribution coming from each term in the above expression. Starting with
\begin{equation}
\label{eq:third_term}
    \begin{aligned}
        &~D\tau\sum_{j=0}^{N-1}\varphi_{0j}\int_{h\ne0}d\boldsymbol{\chi}_h v_a(\boldsymbol{X}_j-\boldsymbol{v}_w t)\sum_{i=0}^{N-1}\boldsymbol{\nabla}_i^2\boldsymbol{\sigma}_j, \\
        =&~D\tau\sum_{j=0}^{N-1}\varphi_{0j}\int_{h\ne0}d\boldsymbol{\chi}_h v_a(\boldsymbol{X}_j-\boldsymbol{v}_w t)\big[\boldsymbol{\nabla}_0^2\boldsymbol{\sigma}_j \\
        &+ \boldsymbol{\sigma}_j\sum_{i\ne0}\boldsymbol{\nabla}_i^2v_a(\boldsymbol{X}_j-\boldsymbol{v}_wt)\big], \\
        =&~\mathcal{O}(\boldsymbol{\nabla}_0^2),
    \end{aligned}
\end{equation}
where, we have used integration by parts from the first to the second step and converted $\boldsymbol{\nabla}_i^2$ to $\boldsymbol{\nabla}_0^2$ using
\begin{equation}
\label{eq:integration_by_parts}
    \begin{aligned}
        \boldsymbol{\nabla}_j~v_a(\boldsymbol{X}_i) &= \partial_{j\beta}~v_a(\sum_{k}\varphi_{ki}\chi_{k\alpha}) = \frac{\varphi_{ji}}{\varphi_{j0}}\partial_{0\beta}~v_a(\boldsymbol{X}_i) \\
        &= \sqrt{N}\varphi_{ji}\partial_{0\beta}~v_a(\boldsymbol{X}_i)\quad\text{as}~\varphi_{0i}=(\sqrt{N})^{-1}~\forall~i,
    \end{aligned}
\end{equation}
where we have switched to index notation for clarity. Thus, this term would not contribute to the flux under the small gradients approximation. Similarly, the contribution from the term with the traveling wave velocity $\boldsymbol{v}_w$ can be determined by plugging the adiabatic expression of $\boldsymbol{\sigma}_j$ as follows
\begin{equation}
\label{eq:first_term}
    \begin{aligned}
        \boldsymbol{\mathcal{I}}_w\equiv&~Dv_w\tau\sum_{j=0}^{N-1}\varphi_{0j}\int_{h\ne0}d\boldsymbol{\chi}_h v_a(\boldsymbol{X}_j-\boldsymbol{v}_w t)\partial_{0\beta}\delta_{\beta,w}\sigma_{j\alpha},\\
        =&~-\frac{\tau^2Dv_w}{d}\sum_{j=0}^{N-1}\varphi_{0j}\int_{h\ne0}d\boldsymbol{\chi}_h v_a(\boldsymbol{X}_j-\boldsymbol{v}_w t) \times\\
        &~\bigg[\sum_{l\ne0}\varphi_{lj}\partial_{0\beta}\delta_{\beta,w}\partial_{l\alpha}[v_a(\boldsymbol{X}_j-\boldsymbol{v}_wt)\varrho] \\
        &-~d\partial_{0\beta}\delta_{\beta,w}\partial_{i\beta}\gamma_i\chi_{i\beta}\sigma_{j\alpha}\bigg] + \mathcal{O}(\partial_{0\alpha}^2),\\
        =&~\frac{\tau^2Dv_w}{d}\sum_{j=0}^{N-1}\varphi_{0j}\int_{h\ne0}d\boldsymbol{\chi}_h [\partial_{l\alpha}v_a(\boldsymbol{X}_j-\boldsymbol{v}_w t)]\times \\
        &~\varphi_{li}\partial_{0\beta}\delta_{\beta,w}[v_a(\boldsymbol{X}_i-\boldsymbol{v}_wt)\varrho] + \tau\gamma_j\mathcal{I}_w + \mathcal{O}(\partial_{0\alpha}^2),\\
        =&~\frac{\tau^2Dv_w}{d}\sum_{j=0}^{N-1}\varphi_{0j}\int_{h\ne0}d\boldsymbol{\chi}_h [\partial_{0\alpha}v_a(\boldsymbol{X}_j-\boldsymbol{v}_w t)]\times \\
        &~\sqrt{N}\varphi_{li}^2\partial_{0\beta}\delta_{\beta,w}[v_a(\boldsymbol{X}_i-\boldsymbol{v}_wt)\varrho] + \tau\gamma_j\mathcal{I}_w + \mathcal{O}(\partial_{0\alpha}^2),\\ 
    \end{aligned}
\end{equation}
where we have used integration by parts to go from the second step to the third, and the identity Eq.~\eqref{eq:integration_by_parts} in the final step. This implies,
\begin{equation}
\label{eq:first_term_extra}
        \boldsymbol{\mathcal{I}}_w(1-\tau\gamma_j) =~ \mathcal{O}(\partial_{0\alpha}^2),\\
\end{equation}
i.e., the contribution $\boldsymbol{\mathcal{I}}_w$ is also at least second order in gradients $\mathcal{O}(\boldsymbol{\nabla}_0^2)$. 

Next, we consider the term with the activity
\begin{equation}
\label{eq:fourth_term}
    \begin{aligned}
        &~-\frac{\tau}{d}\sum_{j=0}^{N-1}\varphi_{0j}\int_{h\ne0}d\boldsymbol{\chi}_h v_a\sum_{l=0}^{N-1}\boldsymbol{\nabla}_l\cdot \big[\varphi_{lj}v_a\varrho\big],\\
        =&~-\frac{\tau}{d}\sum_{j=0}^{N-1}\varphi_{0j}\int_{h\ne0}d\boldsymbol{\chi}_hv_a\partial_{0\alpha}[\varphi_{0i}v_a\varrho]\\
        &~- \frac{\tau}{d}\sum_{j=0}^{N-1}\varphi_{0j}\sum_{l\ne0}\int_{h\ne0}d\boldsymbol{\chi}_hv_a\partial_{l\alpha}[\varphi_{lj}v_a\varrho], \\
        =&~-\frac{\tau}{d}\sum_{j=0}^{N-1}\varphi_{0j}\int_{h\ne0}d\boldsymbol{\chi}_h\big[\varrho~v_a\partial_{0\alpha}v_a + v_a^2\partial_{0\alpha}\varrho] \\
        &~+\frac{\tau}{d}\sum_{j=0}^{N-1}\varphi_{0j}\int_{h\ne0}d\boldsymbol{\chi}_h~\varrho\varphi_{lj}^2\sqrt{N}v_a\partial_{0\alpha}v_a, \\
        =&~-\frac{(N-2)\tau}{2d}\rho_0\partial_{0\alpha}v_a^2(\frac{\boldsymbol{\chi}_0}{\sqrt{N}}) - \frac{\tau}{d}v_a^2(\frac{\boldsymbol{\chi}_0}{\sqrt{N}})\partial_{0\alpha}\rho_0 + \mathcal{O}(\partial_{0\alpha}^2),
    \end{aligned}
\end{equation}
where, we have again used integration by parts to go from the second step to the third, and Eq.~\eqref{eq:integration_by_parts}. We have also made use of the properties of the orthogonalizing matrix $\sum_{j=0}^{N-1}\sum_{l\ne0}\varphi_{lj}^2 = \sum_{l\ne0}\delta_{ll} = N-1$. Furthermore, we have Taylor-expanded the quantity $\partial_{0\alpha}v_a^2(\boldsymbol{X}_i)$ as
\begin{equation}
\label{eq:taylor_expansion}
    \partial_{0\alpha}v_a^2(\boldsymbol{X}_i) = \partial_{0\alpha}v_a^2(\sum_{j=0}^{N-1}\varphi_{ji}\boldsymbol{\chi}_j) = \partial_{0\alpha}v_a^2(\varphi_{0i}\boldsymbol{\chi}_0) + \mathcal{O}(\partial_{0\alpha}^2).
\end{equation}
Finally, we use the adiabatic expression of $\boldsymbol{\sigma}_j$ again to evaluate the contributions coming from the remaining term, which we define as $\sum_{i=0}^{N-1}\mathcal{I}_{i\alpha}$ with
\begin{equation}
\label{eq:second_term}
    \begin{aligned}
        \mathcal{I}_{i\alpha} \equiv&~-\tau\varphi_{0j}\int_{h\ne0}d\boldsymbol{\chi}_h[\partial_{i\beta}v_a]\gamma_i\chi_{i\beta}\sigma_{j\alpha}, \\
        =&~\frac{\tau^2}{d}\varphi_{0j}\int_{h\ne0}d\boldsymbol{\chi}_h[\partial_{i\beta}v_a]\gamma_i\chi_{i\beta}\partial_{l\alpha}[\varphi_{lj}v_a\varrho] \\
        &~-\tau^2\varphi_{0j}\int_{h\ne0}d\boldsymbol{\chi}_h[\partial_{i\beta}v_a]\gamma_i\chi_{i\beta}\partial_{l\gamma}[\gamma_l\chi_{l\gamma}\sigma_{j\alpha}] + \mathcal{O}(\partial_{0\alpha}^2),\\
        =&~-\frac{\tau^2}{d}\varphi_{0j}\int_{h\ne0}d\boldsymbol{\chi}_h[\partial_{i\beta}v_a]\gamma_i[\partial_{l\alpha}\chi_{i\beta}]\varphi_{lj}v_a\varrho \\
        &~+\tau^2\varphi_{0j}\int_{h\ne0}d\boldsymbol{\chi}_h[\partial_{i\beta}v_a]\gamma_i[\partial_{l\gamma}\chi_{i\beta}]\gamma_l\chi_{l\gamma}\sigma_{j\alpha} + \mathcal{O}(\partial_{0\alpha}^2),\\
        =&~-\frac{\tau^2}{d}\varphi_{0j}\int_{h\ne0}d\boldsymbol{\chi}_h[\partial_{i\beta}v_a]\gamma_i\varphi_{ij}v_a\varrho -\tau\gamma_i\mathcal{I}_{i\alpha} + \mathcal{O}(\partial_{0\alpha}^2),\\
        =&~-\frac{\tau^2}{d}\varphi_{0j}\int_{h\ne0}d\boldsymbol{\chi}_h\sqrt{N}\varphi_{ij}^2\gamma_i\varrho v_a[\partial_{0\alpha}v_a]\\
        &~-\tau\gamma_i\mathcal{I}_{i\alpha} + \mathcal{O}(\partial_{0\alpha}^2),\\
        =&~-\frac{\tau^2}{2d}\gamma_i\rho_0\partial_{0\alpha}v_a(\frac{\boldsymbol{\chi}_0}{\sqrt{N}})-\tau\gamma_i\mathcal{I}_{i\alpha} + \mathcal{O}(\partial_{0\alpha}^2), 
    \end{aligned}
\end{equation}
where, we have used integration by parts in the first line and in going from second to the third step. We have also used the identity Eq.~\eqref{eq:integration_by_parts} and the Taylor expansion Eq.~\eqref{eq:taylor_expansion} in the penultimate step. Solving the resulting linear equation for $\mathcal{I}_{i\alpha}$, we get
\begin{equation}
\label{eq:second_term_final}
    \sum_{i=0}^{N-1}\mathcal{I}_{i\alpha} = -\frac{\tau}{2d}[N-2+\epsilon]\rho_0\partial_{0\alpha}v_a(\frac{\boldsymbol{\chi}_0}{\sqrt{N}}) + \mathcal{O}(\partial_{0\alpha}^2),
\end{equation}
where $\epsilon$ is the tactic response parameter defined in the main text in Eq.~\eqref{eq:epsilon}. Putting everything together from Eqs.~\eqref{eq:third_term}-~\eqref{eq:second_term_final} in the probability current for $\rho_0$, we have
\begin{equation}
\label{eq:probability_current_final}
    \begin{aligned}
    \boldsymbol{\mathcal{J}} =&~-\frac{\tau\epsilon}{2d}\rho_0\boldsymbol{\nabla}_0v_a^2(\frac{\boldsymbol{\chi}_0}{\sqrt{N}}) - \bigg[D+\frac{\tau}{d}v_a^2(\frac{\boldsymbol{\chi}_0}{\sqrt{N}})\bigg]\boldsymbol{\nabla}_0\rho_0\\
    &~- \boldsymbol{v}_w\rho_0 + \mathcal{O}(\boldsymbol{\nabla}_0^2).
    \end{aligned}
\end{equation}
Finally, we apply the chain rule to write the drift-diffusion equation Eq.~\eqref{eq:drift_diffusion_eq} with the effect drift and diffusivity defined in Eq.~\eqref{eq:effective_drift_diffusivity}.

We also show here that the choice of our simulation parameters satisfy the small gradients approximation in the co-moving frame of the traveling wave. For this, we compare the magnitude of the activity field $v_a$ with i) the magnitude of the gradient of the activity field over the persistence length, $l_p\nabla v_a$, and ii) the magnitude of the gradient of the activity field over the typical bond length of the polymer, $l_b\nabla v_a$. For small gradients approximation to hold, we require $\frac{l_p\nabla v_a}{v_a}\ll1$, as well as, $ \frac{l_b\nabla v_a}{v_a}\ll1$. We have for Fig.~\ref{fig:steady_state_density_waves_length} and Fig.~\ref{fig:average_drift_length}: $\frac{l_p\nabla v_a}{v_a}\sim~10^{-1}$ and $\frac{l_b\nabla v_a}{v_a}\sim~10^{-2}$, and for Fig.~\ref{fig:steady_state_density_waves_topology} and Fig.~\ref{fig:average_drift_topology}: $\frac{l_p\nabla v_a}{v_a}\sim~10^{-1}$ and $\frac{l_b\nabla v_a}{v_a}\sim~10^{-1}$, all of which lie within the limits of validity of the small gradients approximation.

\section{Steady State Density Profile in Co-moving Frame}
\label{sec:appendix_c}
Here, we sketch the derivation to obtain the steady state density profile with a constant flux $J$ (Eq.~\eqref{eq:steady_state_density_co_moving_frame}) in a co-moving box of length $L$ endowed with periodic boundary conditions. We derive everything in one dimension for simplicity, however, it can be easily generalized to higher dimensions. We start with the continuity equation,
\begin{equation}
\label{eq:appendix_c_1}
    \partial_x\rho(x) + b(x)\rho(x) = -D^{-1}(x)J,
\end{equation}
where we defined the function $b(x) = -V(x)/D(x)$ for brevity. Solving this equation with the integrating factor
\begin{equation}
\label{eq:appendix_c_2}
    \begin{aligned}
    \partial_x\bigg[\rho(x)&\exp{\int_0^xdx'~b(x')}\bigg] =\\
    &-JD^{-1}(x)\exp\{\int_0^xdx'~b(x')\}.
    \end{aligned}
\end{equation}
Integrating this equation over a box length ($y$ to $y+L$) and using the periodic boundary conditions, i.e., $\rho(y+L)=\rho(y)$, we have
\begin{equation}
\label{eq:appendix_c_3}
    \begin{aligned}
    \rho(y)\exp&\big[\int_{0}^{y}dx'~b(x')\big]\bigg\{\exp\bigg[\int_{y}^{y+L}dx'~b(x')\bigg]-1\bigg\}\\ &= -J\int_{y}^{y+L}dx~D^{-1}(x)\exp{\big[\int_{0}^{x}dx'~b(x')\big]}.
    \end{aligned}
\end{equation}
Since $D^{-1}(x)$ and $b(x)$ is also periodic, we have $\exp{\int_{y}^{y+L}dx~b(x)} = \exp{\int_{0}^{L}dx~b(x)}$. Rearranging some terms, we arrive at
\begin{equation}
\label{eq:appendix_c_4}
    \begin{aligned}
    \rho(y)\bigg\{1-&\exp\bigg[\int_{0}^{L}dx'~b(x')\bigg]\bigg\}\\
    &= JD^{-1}(y)\int_{y}^{y+L}dx~\exp{\int_{y}^{x}dx'~b(x')}.
    \end{aligned}
\end{equation}
Finally, using $\int_{0}^{L}dy~\rho(y) = \rho_bL$, we have
\begin{equation}
\label{eq:appendix_c_5}
    J = \dfrac{\rho_bL\bigg\{1-\exp\bigg[\int_{0}^{L}dx'~b(x')\bigg]\bigg\}}{\int_{0}^{L}dy~D^{-1}(y)\int_{0}^{L}dx~\exp\bigg[\int_{y}^{y+x}dx'~b(x')\bigg]}.
\end{equation}
Plugging this expression for $J$ in Eq.~\eqref{eq:appendix_c_4}, we get the expression for the steady state density profile
\begin{equation}
\label{eq:appendix_c_6}
    \frac{\rho(y)}{\rho_b} = \dfrac{LD^{-1}(y)\int_{0}^{L}dx~\exp{\int_{y}^{y+x}dx'~b(x')}}{\int_{0}^{L}dy~D^{-1}(y)\int_{0}^{L}dx~\exp{\int_{y}^{y+x}dx'~b(x')}}.
\end{equation}
\section{Simulation Details}
\label{sec:appendix_d}
All numerical data has been generated using Langevin dynamics simulations of the monomer equations of motion in Eq.~\eqref{eq:monomer_eom} and their orientations in Eq.~\eqref{eq:orientation_eom}. These equations are discretized using the standard Euler-Maruyama scheme and read
\begin{equation}
\label{eq:monomer_eom_discretised}
    \begin{aligned}
        \boldsymbol{X}_i(t+\Delta t) =& ~\boldsymbol{X}_i(t) + \sqrt{2\Delta tD}\boldsymbol{\xi}_i(t) \\
        +&~\mu\Delta t\bigg[-\sum_{j=0}^{N-1}M_{ij}\boldsymbol{X}_j(t) + v_a(\boldsymbol{X}_i(t) - \boldsymbol{v}_wt)\boldsymbol{\eta}_i(t)\bigg], \\
        \boldsymbol{\eta}_{i}(t+\Delta t) =& ~\boldsymbol{\eta}_i(t) - \tau^{-1}\Delta t\boldsymbol{\eta}_i(t) + \sqrt{2\Delta t(\tau d)^{-1}}\boldsymbol{\zeta}_i(t),
    \end{aligned}
\end{equation}
where $\Delta t$ is the discretization timestep, and was chosen as $\Delta t = 0.01$ for all simulations. $\boldsymbol{\xi}_i(t)$ and $\boldsymbol{\zeta}_i(t)$ are independent Gaussian random variables with zero mean and unit variance. All simulations have been performed in a box of size $L=10$ with periodic boundary conditions. The steady state density profiles in Fig.~\ref{fig:steady_state_density_waves_length} and Fig.~\ref{fig:steady_state_density_waves_topology} were obtained using time-averaging from a single simulation of $10^{9}$ simulation steps. For Fig.~\ref{fig:average_drift_length} and Fig.~\ref{fig:average_drift_topology}, the drift of the center of mass was measured for $10^{3}$ independent trajectories, each $10^{6}$ timesteps long. The numerical errors are within the size of the symbols in all figures.
\end{document}